# Decentralization in Bitcoin and Ethereum Networks

Adem Efe Gencer[1,2], Soumya Basu[1,2], Ittay Eyal[1,3], Robbert van Renesse[1,2], and Emin Gün Sirer[1,2]

[1] Initiative for Cryptocurrencies and Contracts (IC3)
[2] Computer Science Department, Cornell University
[3] Electrical Engineering Department, Technion

**Abstract.** Blockchain-based cryptocurrencies have demonstrated how to securely implement traditionally centralized systems, such as currencies, in a decentralized fashion. However, there have been few measurement studies on the level of decentralization they achieve in practice.
We present a measurement study on various decentralization metrics of two of the leading cryptocurrencies with the largest market capitalization and user base, Bitcoin and Ethereum. We investigate the extent of decentralization by measuring the network resources of nodes and the interconnection among them, the protocol requirements affecting the operation of nodes, and the robustness of the two systems against attacks. In particular, we adapted existing internet measurement techniques and used the Falcon Relay Network as a novel measurement tool to obtain our data. We discovered that neither Bitcoin nor Ethereum has strictly better properties than the other. We also provide concrete suggestions for improving both systems.

## 1 Introduction

Cryptocurrencies are emerging as a new asset class, with a market capitalization of about $150B as of Sept 2017 [15], a growing ecosystem, and a diverse community. The most prominent platforms that account for over 70% of this market are Bitcoin [57] and Ethereum [28, 70]. The underlying technology, the *blockchain*, achieves consensus in a decentralized, open system and enables innovation in industries that conventionally relied upon trusted authorities. Some examples of such services include land record management [3], domain name registration [51], and voting [55]. The key feature that empowers such services and makes these platforms interesting is *decentralization*. Without it, such services are technologically easy to construct but require trust in a centralized administrator.

Decentralization is a property regarding the fragmentation of control over the protocol. In the Bitcoin and Ethereum protocols, users submit transactions for miners to sequence into blocks. Better decentralization of miners means higher resistance against censorship of individual transactions. For communication, Bitcoin and Ethereum also have a peer-to-peer network for disseminating block and transaction information. Both Bitcoin and Ethereum also contain *full nodes*, which serve two critical roles: (1) to relay blocks and transactions to miners (2) and to answer queries for end users about the state of the blockchain. Understanding the network properties of full nodes is crucial for protocol design and analysis of each network's resilience to attacks. Ongoing research explores ways to make the Bitcoin and Ethereum networks more decentralized without measurements on the underlying network. Hence, debates and decisions about the underlying networks are often based on assumptions rather than measurement.

In this paper, we present a comprehensive measurement study on decentralization metrics in these operational systems and shed light on whether or not existing assumptions are satisfied in practice. We adapt prior Internet measurement techniques for Bitcoin and Ethereum and use novel approaches to obtain application layer data. Our main data sources are (1) direct measurements of these networks from multiple vantage points, (2) a Bitcoin relay network called *Falcon* that we deployed and operated for a year, and (3) blockchain histories of Bitcoin and Ethereum. Our study presents findings regarding the network properties, impact of protocol requirements, security, and client interactions.

This paper makes three contributions. First, it provides new tools and techniques for measuring blockchain-based cryptocurrency networks. The key tool introduced here is the Falcon relay network that we built to serve as a backbone for ferrying blocks. This network was deployed for Bitcoin across five continents, providing a unique vantage point on pruned blocks. Second, we perform a comparative study of decentralization metrics in Bitcoin and Ethereum. Our key findings are: (1) the Bitcoin network can increase the bandwidth requirements for nodes by a factor of 1.7 and keep the same level of decentralization as 2016, (2) the Bitcoin network is geographically more clustered than Ethereum, with many nodes likely residing in datacenters. (3) Ethereum has lower mining power utilization than Bitcoin and would benefit from a relay network, and (4) small miners experience more volatility in block rewards in Bitcoin than Ethereum.

## 2 Bitcoin and Ethereum

Bitcoin and Ethereum use Nakamoto consensus [5–7, 57, 38] to regulate transaction serialization in their blockchains. While architecturally very similar, these systems differ significantly in terms of their API, abstractions, and wire protocol.

### 2.1 The Bitcoin Protocol

Bitcoin is a protocol that sequences transactions into groups called blocks. The protocol targets a block production interval of 10 minutes with a maximum size of 1 MB. At the time of our measurements, the last 100 blocks had a 0.99 MB median block size and a 9.8 minute mean interval. The wire protocol implements a peer-to-peer network based on flooding block and transaction announcements.

The peer to peer network is formed through point to point links. To form a link, clients establish a TCP connection and perform a protocol-level three-way handshake. The protocol-level handshake exchanges the state of each client, such as the height of the blockchain and a version string associated with the software being run. When a client discovers or receives a new block, it floods the network with the hash of the block. If a neighboring client needs the block, it requests the block based on the hash value. There are many different block formats, such as compact [17] and Merkle [44] blocks, but we focus on retrieval of full blocks.

### 2.2 The Ethereum Protocol

The Ethereum protocol [28] focuses on providing a platform to facilitate building decentralized applications on its blockchain. To sequence transactions, Ethereum adopts a design inspired by Nakamoto consensus and the GHOST protocol [64].

Ethereum adopts a chain selection rule to harness the residual mining power in pruned blocks for improved security. The protocol includes such blocks, called *uncle*s, in its blockchain and rewards the corresponding miners [70]. Ethereum targets a block interval between 10 to 20 seconds [41]. The block size is indirectly determined by an execution fee, called *gas*, that fluctuates over time. At the time of our measurements, the last 100 blocks were generated with a 2.9 KB median block size and a 16.3 second average interval.

Ethereum employs a UDP-based node discovery mechanism inspired by Kademlia [54], but the rest of the P2P communication is over TCP. Unlike Bitcoin, messages between nodes are encrypted and authenticated. Ethereum's wire protocol is poorly documented, so we rely on the client implementations [42, 61, 19, 60] and Ethereum wiki pages [26, 25, 29, 30, 27] for information.

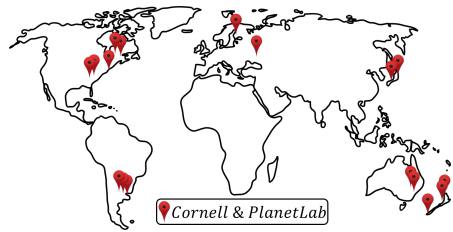

Fig. 1: The measurement infrastructure is built on 18 globally distributed nodes.

In Ethereum, clients request blocks by the corresponding block hash. Older clients request blocks, which consist of a body and header, while newer clients request each piece separately. The measurement system in this paper focuses on retrieval of full blocks and block bodies.

## 3 Measurement Infrastructure

Blockchain-based cryptocurrencies operate on global peer-to-peer networks that span multiple administrative domains. Measurement of such networks concerns the exploration of the relationship between peers, the capabilities of individual peers, and the properties of the system as a whole–e.g. its security and fairness. To characterize Bitcoin and Ethereum, we deployed *Blockchain Measurement System* (*BMS*), a measurement system than ran experiments of varying duration–from a few days up to 12 months.

*Network Properties.* BMS uses multiple vantage points in order to gain a comprehensive view of the cryptocurrency networks. To capture the evolution of these networks, BMS has been continuously collecting data regarding the provisioned bandwidth of peers and peer-to-peer latency. BMS first connects to a peer, collects measurements, and then disconnects before proceeding to the next peer. These measurements target (1) Bitcoin nodes connected over IPv4, IPv6, and Tor [23] and (2) Ethereum nodes connected over IPv4. As of May 2017, Ethereum does not have any Tor nodes mainly because Tor is exclusively TCP, whereas Ethereum node discovery is UDP-based. Moreover, this study excludes Ethereum's IPv6 network because BMS was unable to discover enough nodes to reach generalized conclusions. Table 1 shows the timeline of the data collection for each network and the number of nodes measured in each measurement.

| Measurement | Network | | Num. Nodes | Dates |
|---|---|---|---|---|
| Bandwidth (All) Latency (BTC IPv4) *(Single Beacon)* | BTC | IPv4 | 3441 | Jan 11–16;Jan 30–Mar 16 |
| | | IPv6 | 515 | Jan 13–14;Apr 20–25 |
| | | Tor | 127 | Jan 13;Apr 23–25 |
| | ETH | IPv4 | 285 | Mar 27–Apr 25 |
| Peer-to-Peer Latency *(Mult. Vantage Pts.)* | BTC | IPv4 | 3390 (5.7M edges) | Jan 10–15; Jan 30–Mar 01 |
| | ETH | IPv4 | 4302 (9.3M edges) | Mar 01–Apr 11 |
| Latency *(Single Beacon)* | BTC | IPv6 | 845 | Jan 13–14; Feb 03–Apr 25 |
| Pruned Blocks | BTC | IPv4 | 5977 | May 5 2016 – Apr 29 |

Table 1: Timeline of measurements. All dates are in 2017 unless otherwise noted.

To estimate the peer-to-peer latency, BMS uses multiple vantage points geographically distributed across the world. Figure 1 shows the geographic distribution of the measurement infrastructure. 15 out of 18 nodes reside in PlanetLab's global research network [14] and the remaining three nodes are part of Cornell's academic network, located in Ithaca, NY.

To measure the provisioned bandwidth of nodes in Bitcoin and Ethereum, BMS used nodes with extensive resources. In particular, measuring the maximum bandwidth that Bitcoin and Ethereum nodes have access to requires nodes with (1) high download capacities to ensure that the bottlenecks are not in the measurement apparatus, and (2) sufficient disk capacities to store detailed results. Since these machines need access to orders of magnitude higher bandwidth capacity than what is achievable on shared infrastructure, such as PlanetLab nodes, some BMS data was collected using dedicated, well-provisioned beacon nodes located at Cornell University.

Finally, BMS needs to pick a sample of nodes from the Bitcoin and Ethereum networks. As a sample, BMS uses a list containing nodes from Bitcoin and Ethereum node crawling sites [1, 31], and a locally deployed Ethereum supernode configured with a high peer limit. Interpretations in this paper assume that inferences made from the reachable public nodes are representative of their entire networks. In reality, these networks contains nodes that are not visible to the public, e.g. they are behind a NAT or a firewall. One such class of nodes are part of *mining*. While much of the mining infrastructure is private, prior measurement work shows that mining operations often have gateway nodes to communicate with the peer-to-peer network [56]. The properties of internal mining pool nodes are orthogonal to the focus of this paper.

*Blockchain Information.* A naive approach to obtaining information about the blockchain would be to simply run a Bitcoin and Ethereum node. However, this precludes information that cannot be obtained through the respective wire protocols. Many important decentralization metrics center around the analysis of blocks that are not part of the main blockchain. In Ethereum, many of these blocks become uncles which can simply be requested through the wire protocol.

In Bitcoin, however, a block that is not part of the main blockchain simply becomes *pruned*. Pruned blocks in Bitcoin have no effect on the state of the system, they are deleted by clients without impacting correctness. Thus, it is crucial to connect directly to miners to capture pruned blocks.

A critical component of BMS to observe pruned blocks is the Falcon Relay Network, which relays blocks between Bitcoin miners. The Falcon Relay Network uses cut-through routing to quickly disseminate blocks worldwide, which incentivizes miners to connect to Falcon. Indeed, Falcon is directly connected to at least 36.4% of the entire hashpower in Bitcoin. Since there is just one other operational relay network for Bitcoin [18, 16], Falcon has observed blocks that have not been seen on other well-connected nodes [8].

## 4 Measurements

In this section, we present the measurements taken by BMS. In each measurement, we describe the methodology, followed by the results of our analysis. As with any measurement study of a large-scale, uninstrumentable artifact, measurements are not perfect; we conclude each section by addressing some potential sources of error and their mitigation.

### 4.1 Provisioned Bandwidth

Provisioned bandwidth is an estimate on a node's transmission capacity characterizing how much bandwidth the node has to communicate with the rest of the cryptocurrency network. Greater provisioned bandwidth helps miners to propagate/collect blocks to/from the network faster. Thus, it becomes more difficult for a malicious miner to situate themselves in the network to achieve the rushing property [35] and attack the blockchain. Knowledge of provisioned bandwidth also aids in setting protocol parameters, such as the block size and frequency.

**Methodology.** BMS measures the provisioned bandwidth of each peer by requesting a large amount of data from each peer and seeing how fast the peers can stream the data to BMS's measurement nodes. BMS does this by asking for blocks that were first seen over a year ago – similar to how a stale node asks for blocks to sync state. Each request asks for the same set of blocks in Bitcoin and blocks or the corresponding bodies in Ethereum. Next, BMS divides the time into epochs and records the number of bytes received during each epoch. This process continues until either BMS receives all data or a predefined timeout of 30 seconds is reached. A long timeout helps BMS eliminate effects from TCP slow start and other initialization noise as well as identify and eliminate spurious spikes in throughput caused by buffering in the kernel by BMS. Finally, BMS processes the collected data to determine the provisioned bandwidth. To do so, first, it identifies the independent data streams by combining successive epochs containing active data transfers. Then, it eliminates streams that are shorter than 500 milliseconds to mitigate initialization artifacts such as TCP slow start. BMS then outputs the maximum observed throughput among the remaining distinct continuous streams as the provisioned bandwidth of the remote peer.

The experiments in this paper are run on servers with 1 Gbps links at Cornell University. This has not changed from 2016 to 2017, which allows us to make comparisons to a previous study in 2016 [20].

|          | **Bitcoin** | | | **Eth.** |
|          | IPv4 [Mbps] | IPv6 [Mbps] | Tor [Mbps] | IPv4 [Mbps] |
|---|---|---|---|---|
| 10%      | 5.7   | 11.0  | 2.1 | 3.4   |
| 33%      | 23.3  | 45.2  | 3.1 | 11.2  |
| 50%      | 56.1  | 78.2  | 4.1 | 29.4  |
| 67%      | 91.1  | 94.3  | 5.6 | 68.3  |
| 90%      | 177.0 | 207.9 | 8.1 | 144.4 |
| Avg.     | 73.1  | 86.5  | 4.7 | 55.0  |
| Std. Dev.| 68.4  | 66.9  | 2.4 | 58.8  |

(a) Provisioned bandwidth statistics.

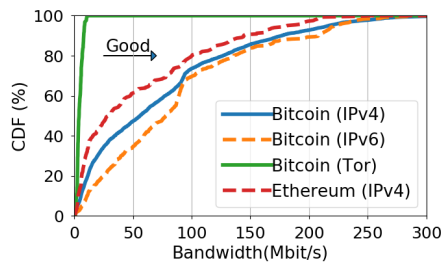

(b) CDF

Fig. 2: Statistics on distribution of provisioned bandwidth and the CDF.

**Results.** Table 2a summarizes per-node bandwidth results that BMS has collected. We see that Bitcoin nodes in both IPv4 and IPv6 networks have consistently higher bandwidth compared to Ethereum IPv4 nodes. In particular, the median Bitcoin IPv4 and IPv6 nodes have about $1.9\times$ and $2.7\times$ the bandwidth of the typical Ethereum IPv4 node. In contrast, Bitcoin Tor nodes have an order of magnitude lower bandwidth compared to directly connected nodes, though they are not unusable – e.g. 90% has more than 2 Mbit/s. Ongoing research explores alternatives to the Tor network that also provide efficient communication [50].

Figure 2b shows the cumulative distribution of the bandwidth measurements. Steep increases in the Bitcoin IPv4/IPv6 curves at around 10 Mbps and 100 Mbps regions represent typical bandwidth capacities of a home user, and a typical Amazon EC2 Bitcoin instance. For Ethereum, we observe a similar accumulation around 10 Mbps region, but the bandwidth is more evenly distributed over the remaining nodes. As the long tailed distribution and higher standard deviation indicates, bandwidth of Bitcoin IPv4/IPv6 nodes are spread out over a wider range of values compared to Ethereum nodes. While the most well provisioned Bitcoin nodes have around 300 Mbps of spare bandwidth, the highest Ethereum node capacity that BMS has observed is limited to 250 Mbps.

One of the most interesting discoveries of this study is that the Bitcoin network has improved tremendously in terms of its provisioned bandwidth. The results show that Bitcoin IPv4 nodes, which used to be connected to the network with a median bandwidth of 33 Mbit/s in 2016 [20], have a median bandwidth of 56 Mbit/s, as of February 2017. In other words, the provisioned bandwidth of a typical full node is now $1.7\times$ of what it was in 2016.

Critical system parameters, such as the maximum block size and block frequency, can be increased when the provisioned bandwidth increases. The increase in provisioned bandwidth suggests that the block size can be increased by a factor of 1.7 without increasing centralization beyond its de facto level in 2016.

**Caveats.** As with every measurement technique in the real world, our results above are subject to experimental limitations and expected errors. The accuracy of the measurements may drop under certain circumstances, including the cases where: (1) the network bottleneck lies on the side of the measurement beacon rather than the remote peer, (2) network traffic on the side of BMS interferes with the collected results, (3) the remote peer intentionally shapes the traffic to selectively limit the bandwidth available to BMS, for instance via bandwidth throttling, and (4) different steady state bandwidth between Bitcoin and Ethereum, skewing the numbers for one system over another The setup of our bandwidth infrastructure helps minimize potential inaccuracies due to the first two issues. Moreover, analysis of popular Bitcoin [5] and Ethereum client implementations [42, 61, 19, 60] shows that the third case is not supported by this software and would require additional, potentially non-trivial, work to set up. To verify the impact of the last issue, we ran an Ethereum and Bitcoin client and saw that their bandwidth consumption differed by 0.2 Mbps, which introduces about a 1% error on our measurements above.

In addition to our analysis above, we also expect to see certain artifacts in our data. As noted above, we see clusters of nodes around 10 Mbps and 100 Mbps, which are typical bandwidth capacities of home and EC2 users, respectively.

### 4.2 Network Structure

The structure of the peer-to-peer network impacts the security and performance for cryptocurrencies. A geographically clustered network can quickly propagate a new block to many other nodes. This makes it more difficult for a malicious miner to propagate conflicting blocks/transactions quicker than honest nodes. However, a less clustered network may mean that full nodes are being run by a wider variety of users which is also good for decentralization.

**Methodology.** Since it is not possible to obtain direct measurements between peers we do not control, we use the state of the art estimation techniques to establish bounds and gain insights into network structure.

*Single Beacon Latency.* We first collect direct ICMP ping measurements from BMS nodes to all peers in the network. We report the minimum observed ping latency, as it provides a physical bound on the distance to the BMS beacon.

*Peer-to-Peer Latency.* Measuring the peer-to-peer latency requires access to the end points. In both Bitcoin and Ethereum, peers do not reveal their neighbors. Hiding the network structure boosts privacy and security [45, 56], but also makes it harder to infer properties about the network. BMS provides latency estimates for a superset of the actual links between known peers. We do not normalize for the slightly different network sizes, 3390 for Bitcoin and 4302 for Ethereum, as our samples from both networks were very similar. Since measuring peer-to-peer latencies directly is not feasible, we establish bounds from observed latencies from multiple beacons, using techniques from prior literature [37]. BMS starts with the measurements taken from a single beacon. Then, it uses the triangle inequality to estimate the upper and lower bounds for the latency between peers. Repeating this process from other vantage points yields a set of bounds for each pair of peers. Finally, BMS determines a range for latency estimates between each peer by picking the maximum lower bound and the minimum upper bound. The paper also presents the average of the lower bound and upper bound latency between peers. In this study, BMS includes nodes that do not support the DAO fork [10] in its measurements for Ethereum.

*Geographical Distance.* BMS takes the minimum of repeated latency measurements to eliminate transient network effects and capture the geographic distance between two nodes [43, 13, 69]. BMS also uses IP geolocation data to calculate distances between peer nodes as an additional validation on our results. To calculate distances, BMS applies the Haversine formula [63] using the coordinate values gathered from an IP-based geolocation service [46].

**Results.** Our measurements indicate considerable differences between P2P latencies of Bitcoin and Ethereum IPv4 networks, summarized in Table 2 and PDF graphed in Figure 3.

We find that Bitcoin has many more nodes that are closer geographically than Ethereum. Figure 3 shows that Ethereum's most likely latencies are centered around 120ms, while Bitcoin nodes tend to be clustered around 50ms. Only 13% of Ethereum latencies are under 100ms, while Bitcoin has a surprisingly high 46%. Additionally, the estimated peer-to-peer latency between Ethereum nodes is 26.7% higher than Bitcoin on average. This geographic proximity between nodes, along with the observation that Bitcoin has many nodes with 100 Mbps of provisioned bandwidth (see Section 4.1), seems to indicate that many Bitcoin nodes are run in datacenters. 56% of Bitcoin's nodes and 28% of Ethereum's nodes belong to an autonomous system that provides dedicated hosting services, a difference significant at the 1% significance level.

|  | Single Beacon Bitcoin | | Peer-to-Peer Bitcoin | Eth. |
|---|---|---|---|---|
|  | IPv4 [ms] | IPv6 [ms] | IPv4 [ms] | IPv4 [ms] |
| 10% | 29 | 40 | 48 | 92 |
| 33% | 78 | 80 | 79 | 125 |
| 50% | 89 | 95 | 109 | 152 |
| 67% | 98 | 95 | 152 | 200 |
| 90% | 201 | 165 | 286 | 276 |
| Avg. | 97 | 103 | 135 | 171 |
| Std. Dev. | 59 | 62 | 88 | 76 |

Table 2: Min single beacon latencies observed and P2P latency estimates.

Indeed Ethereum nodes are not accumulated in a single geographical region, but are more evenly distributed around the world. Figure 3c shows the CDF of distances between peer to peer nodes based on IP geolocation information. The results corroborate our findings based on network latency measurements and show that Ethereum nodes are geographically further apart than Bitcoin. As additional evidence, when we use geolocation on the P2P distances and plot the CDF in Figure 3c, we see that Ethereum nodes are further apart than Bitcoin.

**Sanity Checks.** The first two columns of Table 2 present single beacon latency in Bitcoin IPv4/IPv6 networks. The results indicate that both the median and the average latency to IPv4 nodes are smaller than IPv6 nodes. As there are fewer IPv6 nodes than IPv4 nodes, we expect this result since IPv4 nodes are more likely to be closer to our beacons.

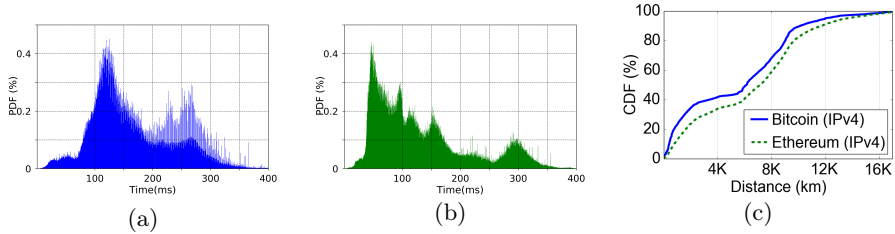

Fig. 3: The histogram of P2P latencies in Ethereum (Fig 3a) and Bitcoin (Fig 3b), as well as the CDF of geographical distances (Fig 3c).

While there has been a large body of work showing the prevalence of triangle inequality violations in the Internet [52, 67, 12], there are several reasons BMS's measurements are not affected significantly. First, such violations were shown to occur less than 5% of network snapshots [52]. Since we take the minimum latency observed from a beacon, triangle inequality violations will only occur in our dataset less than 1% of the time [52]. TIVs are also significantly less prevalent when dealing with latencies less than 300 ms, which includes almost our entire dataset [67]. To ensure that the above results hold for our dataset as well, we used a geolocation service as ground truth to verify our results.

One other limitation in our study is that it is impossible to collect measurements using ICMP pings from nodes that block ICMP traffic and from Tor nodes that only communicate over TCP.

### 4.3 Distribution of Mining Power

Mining on cryptocurrency networks is a complex process that typically requires large computation power. With the current mining difficulty of Bitcoin and Ethereum, using commodity hardware to generate blocks is not feasible [21] which centralizes the mining process somewhat. However, as long as there are many different entities mining, the system is still decentralized. We compare the decentralization of the mining process between Bitcoin and Ethereum.

**Methodology.** To identify the power of miners in Bitcoin and Ethereum, we examined their weekly distribution over the last 10 months starting on July 15, 2016. Our mining power estimations are based on the ratio of main chain blocks generated by distinct entities. Hence, pruned blocks in Bitcoin and uncles in Ethereum do not affect these estimations. In both networks, miners voluntarily disclose their identity as part of each block they mine. We gathered this data from a public API for Bitcoin [9] and a blockchain explorer for Ethereum [32]. In Bitcoin, 1.8% of the blocks were unidentified, which we treated as if they were generated by distinct individual miners. Finally, we manually processed identities to detect and merge duplicates. This includes pools operated by the same administrator [47] and multiple identities representing the same pool, which we identified by looking for the same pool name with a corresponding tag, e.g. 'DwarfPool1' and 'DwarfPool2'. While it is important to note that miners can be either solo miners or mining pools, this distinction is immaterial for the purposes of this analysis. The argument that mining pools provide a degree of decentralization due to mining pool participants having a check on pool operator behavior has no empirical support. For instance, censorship attacks by pool operators are are difficult, if not impossible, to detect by pool participants. Additionally, when miners exceeded the 51% threshold on three separate occasions in Bitcoin's history, the pool participants did not disband the pool despite clear evidence of a behavior widely understood to be unacceptable. Most crucially, whether mining pools provide a degree of decentralization is inconsequential for the purposes of this paper, which provides an accurate historical account. We report what happened at the time the blocks were mined, as recorded on the blockchain. As such, it is immaterial whether the miners were part of a pool or whether they were solo miners. At the time a block was committed to the chain, pool participants were plaintively cooperating as part of the same mining entity.

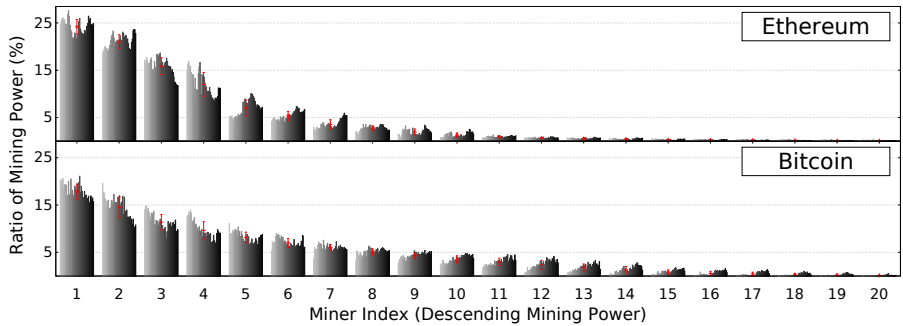

Fig. 4: Distribution of mining power in Bitcoin and Ethereum networks. Bars indicate observed standard deviation from the average.

**Results.** For each week of the analysis period, we calculated the corresponding mining power of entities and ranked each miner accordingly. Figure 4 shows the top 20 weekly mining power distribution in the Ethereum and Bitcoin networks. Each group of bars represents a chronologically ordered collection of weekly *mining power ratios*, defined as the fraction of blocks contributed by a miner.

Figure 4 illustrates that, in Bitcoin, the weekly mining power of a single entity has never exceeded 21% of the overall power. In contrast, the top Ethereum miner has never had less than 21% of the mining power. Moreover, the top four Bitcoin miners have more than 53% of the average mining power. On average, 61% of the weekly power was shared by only three Ethereum miners. These observations suggest a slightly more centralized mining process in Ethereum.

Although miners do change ranks over the observation period, each spot is only contested by a few miners. In particular, only two Bitcoin and three Ethereum miners ever held the top rank. The same mining pool has been at the top rank for 29% of the time in Bitcoin and 14% of the time in Ethereum. Over 50% of the mining power has exclusively been shared by eight miners in Bitcoin and five miners in Ethereum throughout the observed period. Even 90% of the mining power seems to be controlled by only 16 miners in Bitcoin and only 11 miners in Ethereum. Hence, both platforms rely heavily on very few distinct mining entities to maintain the blockchain. Indeed, we see in Figure 5 that the mining power trends can be fit as exponential distributions with curves $0.21e^{-0.19x}$ and $0.35e^{-0.30x}$ in Bitcoin and Ethereum, respectively. These curves yield a coefficient of determination value of 0.99.

These results show that a Byzantine quorum system [53] of size 20 could achieve better decentralization than proof-of-work mining at a much lower resource cost. This shows that further research is necessary to create a permissionless consensus protocol without such a high degree of centralization.

**Sanity Checks.** Similar to other works in the literature [58, 68], we assume that miners accurately self-identify themselves. A miner that contributes a significant portion of the hash power to the cryptocurrency can exert some amount of influence over protocol changes. Thus, it is likely that miners will want to claim blocks that they generated as their own.

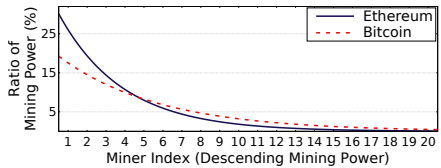

Fig. 5: Exponential trendlines for the average distribution of mining power.

While strong miners gain political clout and attract more members, getting too large raises alarms among the community about centralization. Thus, such miners may conceal or obfuscate this information to appear less powerful – e.g. by generating multiple identities. For instance, two major mining pools, Ethpool and Ethermine, publicly reveal that they share the same admin [47]. Thus, any analysis based on the voluntary miner data skews toward a more decentralized network than the reality.

### 4.4 Mining Power Utilization

Mining power utilization [34], which measures the fraction of mined blocks that remain in the main chain, is a metric for evaluating the efficiency of a protocol, as well as a second order metric for robustness against rollbacks. As mining power utilization increases, the protocol is able to convert more of the energy spent to useful work, and therefore the cost to launch an attack is higher.

**Methodology.** To study the mining power utilization, we analyzed weekly and daily distribution of pruned blocks in Bitcoin and uncles in Ethereum, compared to the main chain blocks. We retrieved this data from (1) the Falcon network, (2) a local Bitcoin client, and (3) public blockchain explorers for Bitcoin [9] and Ethereum [32]. In particular, the Bitcoin blockchain explorer and Falcon exclusively provided 12% and 20% of the total 124 pruned blocks, respectively. Both sources commonly discovered the remaining 68%.

**Results.** Figure 6a and Figure 6b show weekly and daily distributions of mining power utilization in Bitcoin and Ethereum networks, respectively. The results show that Bitcoin utilization is always above 99%, which means that a pruned block in Bitcoin is a relatively rare event. In contrast, daily utilization in Ethereum is typically between 90% to 94% range and never goes above the 97% threshold. During 2016, Ethereum faces occasional drops in its utilization down from 74% to 88%, including (1) the days following the exploitation of the DAO vulnerability [10] from June 17 to 18, (2) attacks on Ethereum network [11, 66] between September 22 to October 19, and (3) the days following the Spurious Dragon hard fork [48] between November 23 to 29. These results indicate a strong relationship between mining power utilization and real life events in Ethereum. This may be the result of preventive measures that spam the network to slow down the DAO attacker, bad actors generating blocks with excessive resource demands, and miners with outdated clients. These results indicate that a relay network, like Falcon, would be greatly beneficial to the Ethereum network.

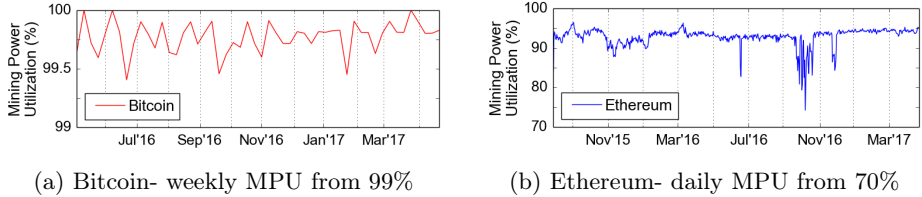

(a) Bitcoin- weekly MPU from 99%    (b) Ethereum- daily MPU from 70%

Fig. 6: Mining power utilization (MPU) for Bitcoin and Ethereum

**Sanity Checks.** The design of the Ethereum protocol requires peers to store and propagate uncle blocks, which are not on the main chain. In contrast, Bitcoin's blockchain only stores the main chain so peers do not propagate pruned blocks. Hence, capturing such blocks in Bitcoin requires actively watching the network. While the Falcon relay network provides a strong incentive to miners to relay blocks through it, some miners may choose not to do so. Consequently, we may be missing some pruned blocks that were generated by the Bitcoin network.

### 4.5 Fairness

Section 4.3 presented the mining power distribution, which looks at the main chain presence of miners. The impact of this distribution on a miner's pruned block rate is unclear. To study this relationship, we examine *fairness* defined as the ratio of a miner's share of pruned blocks to her mining power. In a fair protocol, min-ers generate pruned blocks proportional to their mining power; hence, the fairness is close to 1. A fairness greater than 1 implies that the miner is at a disadvantage, while a fairness less than 1 implies that the miner has an advantage.

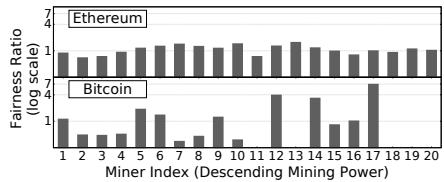

Fig. 7: Fairness distribution. Zero fairness means no pruned block from miner.

**Methodology.** We used the Falcon network, and a Bitcoin blockchain explorer [8] to retrieve pruned Bitcoin blocks. These sources have, respectively, provided 109 and 99 blocks, yielding a total of 124 distinct pruned blocks. We collected uncles from an Ethereum blockchain explorer [32].

Similar to Section 4.3, our results here also assume that miners voluntarily identify themselves in uncles/pruned blocks. Another caveat here lies in gathering pruned blocks. While we incentivize miners to relay blocks through Falcon, there is no guarantee that they necessarily will do so. We suspect that explicit storage of uncles in Ethereum captures a larger proportion of pruned blocks.

**Results.** Figure 7 shows the distribution of fairness of 20 miners with the highest mining power. The results indicate that, in both networks, the top four miners generally are more successful at appending blocks to the main chain. We run the Kolmogorov-Smirnov goodness of fit test with a p-value of 0.01 to compare the fairness distributions of Bitcoin and Ethereum. Perhaps surprisingly, we see that the fairness of Ethereum and Bitcoin differ significantly from each other keeping a constant time period. The reason for this difference is a much larger standard deviation in Bitcoin's miner fairness compared to Ethereum (1.72 versus 0.25). The mean of both fairness distributions, however, are very similar, with Ethereum at 1.08 and Bitcoin at 1.22.

A high variance results in centralization pressure since smaller miners will have a more difficult time affording the loss of revenue due to a transiently high fairness score. This high variance is a direct result of a significantly smaller number of blocks being generated in Bitcoin. Since Ethereum has a higher block frequency than Bitcoin, smaller miners have a more predictable payoff than larger miners. This makes Ethereum more predictable to mine for smaller miners due to the lower variance in block rewards. Thus, it is important for blockchain protocols to take variance of the block rewards in addition to the mean.

Simply increasing the block frequency may not be the solution to decrease the variance of block rewards since the mining power distribution may be affected as well. The increased block frequency in Ethereum may be part of the cause of the slightly more centralized mining power distribution (see Section 4.3).

**Sanity Checks.** Similar to Section 4.4, our results here also assume that miners voluntarily identify themselves in uncles/pruned blocks. As before, if the miners are lying, they are likely to present a more fair system than reality. Another caveat here lies in gathering pruned blocks. While we incentivize miners to relay blocks through Falcon, there is no guarantee that they will. We suspect that explicit storage of uncles in Ethereum allows for more accurate analysis.

Finally, Bitcoin has a significantly lower block generation frequency than Ethereum. On top of that, Bitcoin also has a lower pruned block rate than Ethereum does, which means it has significantly fewer pruned blocks. Thus, this fairness metric is much noisier in Bitcoin compared to Ethereum.

## 5 Related Work

Network measurements in blockchain-based systems have mainly focused on Bitcoin. One such study [22] demonstrated that the latency is the dominating factor in propagation of blocks smaller than 20 KB. Following work [20] has shown that (1) this limit has increased to 80 KB and (2) nodes are provisioned with substantially higher bandwidth capacity than what the protocol demands. Feld et al. [36] pointed out a strong AS-level centralization that may impact Bitcoin network's connectivity – i.e. 10 ASes contain over 30% of peers. Recent work [2] presented the level of vulnerability, showing that 13 ASes cover the same fraction of peers, but only 39 IP prefixes host half of the overall mining power. Ours is the first work that does a similar type of study on Ethereum as well.

Other work studied various aspects of the Bitcoin overlay network. Miller et al. [56] found that a small fraction of the network, containing around 100 nodes, represents more than 75% of the mining power. The study conjectured that these nodes are well-connected to major mining pools; hence, provide higher efficiency in broadcasting blocks. Biryukov et al. [4] examined how peer neighbors discover IP addresses that correspond to pseudonymous identities. Another study [49] deanonymized peers by observing anomalous relaying behavior in network. Pappalardo et al. [59] observed that low value transactions may experience waiting times of over a month. Other work measured churn and geolocated peers [24]. Gervais et al. [40] discussed centralization concerns regarding the client development process, distribution of mining power, and spendable coins. Most of these works focus on attacks and the structure of the overlay network, while this work focuses on the resource capabilities of the nodes used in the overlay network.

Recent work presented ways to reduce resource requirements to participate in blockchain systems. Such solutions enhance decentralization by increasing the diversity of participants. Aspen [39] achieves this through sharding the blockchain. In this system, users store, process, and propagate only the data that is relevant to them, hence need fewer resources to join the network. Another approach [62] relies on authenticated data structures to reduce load on nodes. Relay networks increase network efficiency through faster block propagation. The first such system [16] achieved this by avoiding full block verification and retransmitting known transactions. Falcon, the source of pruned block data in the Bitcoin network in this paper, relies on cut-through routing for faster block propagation. Finally, FIBRE incorporates cut-through routing with compact blocks [17] and forward error correction over UDP. The novelty in our work was utilizing Falcon data in order to gain insights into transient application layer information.

Blockchain explorers [65, 8, 32, 33] provide a variety of data on cryptocurrency networks, including online blockchain history; statistics on blockchain components, transaction fees, and market value; and node information. While these sources of information are useful to the community, this work scientifically tests whether the intuitions provided by these sources of information indeed hold.

## 6 Conclusion

Decentralization in blockchain-based platforms is a component of the value proposition these systems offer. This work presents a comparative assessment of decentralization in two most popular cryptocurrencies, Bitcoin and Ethereum. To do so, it relies on novel measurement techniques to obtain application layer information using the Falcon Network and the application of well-established internet measurement techniques.

Our observations show that Bitcoin has a higher capacity network than Ethereum,but with more clustered nodes likely in datacenters. We also observe that Bitcoin and Ethereum have fairly centralized mining processes and that further research is needed to decentralize permissionless consensus protocols further. In Ethereum, the block rewards have less variance than Bitcoin's. Finally, Ethereum has a lower mining power utilization than Bitcoin, likely due to the high block frequency.

Further, we see that Bitcoin has undergone tremendous growth and can increase the block size by a factor of 1.7x without any decrease in decentralization compared to 2016. Additionally, our study uncovers that the volatility of mining rewards is an important, but often ignored, metric. Finally, we see that Ethereum would likely benefit from a relay network to increase its mining power utilization.

# 7 Acknowledgements

The authors thank Vitalik Buterin and the anonymous reviewers for their feedback on earlier drafts of this manuscript. Ittay Eyal is supported by the Viterbi Fellowship in the Center for Computer Engineering at the Technion. This work was partially funded and supported by AFOSR grant F9550-16-0250, NSF CSR-1422544, NSF CNS-1601879, NSF CNS-1544613, NSF CCF-1522054, NSF CNS-1518779, NSF CNS-1704615, ONR N00014-16-1-2726, NIST Information Technology Laboratory (60NANB15D327, 70NANB17H181), Facebook, Infosys, and IC3, the Initiative for Cryptocurrencies and Smart Contracts. This material is based upon work supported by the National Science Foundation Graduate Research Fellowship Program under Grant No. DGE-1650441. Any opinions, findings, and conclusions or recommendations expressed in this material are those of the authors and do not necessarily reflect the views of the National Science Foundation.